\documentclass[12pt]{article}       
\usepackage{amssymb}
\usepackage{graphicx}
\begin{document} 
\begin{center}
{\large \bf Elastic scattering and spatial inelastic profiles 
of high energy protons}

\vspace{0.5cm}                   

{\bf I.M. Dremin}

\vspace{0.5cm}                       

        Lebedev Physics Institute, Moscow 119991, Russia

\medskip

    National Research Nuclear University "MEPhI", Moscow 115409, Russia     

\end{center}

\begin{abstract}
The ratio of elastic to total proton cross sections is related to the 
darkness of the spatial profile of inelastic interactions by a single 
parameter in the framework of a simple analytical model. Their
critical values at LHC energies are discussed. Two possible variants
of their asymptotical behavior are described.
\end{abstract}

\medskip


\medskip


\medskip

\section{Introduction}

Studies of elastic scattering of high energy protons lead to several unexpected 
results reviewed, e.g., in Refs \cite{ijmp, ufn17}. Among them, the increase of 
the share of elastic scattering with the energy increase by a factor about 1.5 
from ISR-energies to LHC is the most surprising and yet unexplained phenomenon. 
Up to now it is unclear why inelastic processes become losing their competition 
with elastic scattering. With the help of the unitarity condition this
feature can be formulated in terms of the darkness of the spatial profile of 
inelastic interactions which also increases.

One of the ways to understanding these results lies through the 
detailed analysis of the intriguing shape of the elastic
differential cross section with respect to the transferred momentum. 
Its characteristic fast exponential decrease at comparatively small
transferred momenta in the so-called diffraction cone and subsequent
(dip/bump + slower decrease) structure at higher momenta have been carefully
studied by experimentalists in a wide energy interval and, especially, at LHC 
\cite{totem1, totem2, totem3, totem4, atlas1, atlas2}. 

Many phenomenological models were proposed (see, e.g., recent papers
\cite{kmr, glm, bdh, ncc, fgp, dr15, anis, sel, kfk, kfk1, fms, bsw} and 
references therein) for explanation of these peculiarities. Among them,
the kfk-model \cite{kfk, kfk1} must be especially highlighted. The
analytical expressions for the imaginary and real parts of the
elastic scattering amplitude based on some QCD arguments are presented there
as functions both of the transferred momentum measured in experiment and
of the impact parameter relevant to the spatial view of the process.
It describes successfully experimental data in a wide energy interval
from 20 GeV to 8 TeV in the center-of-mass system choosing the energy 
dependence of the adjustable parameters.

In total, there are 8 such parameters each of which contains the energy
independent terms and those increasing with energy $s$ as $\log \sqrt s$ 
and $\log ^2\sqrt s$ (see Eqs (29)-(36) in \cite{kfk}). Thus 8
coefficients should be determined from comparison with experiment at a given
energy and 24 for the description of the energy dependence in a chosen interval. 
Besides, there is another constant $a_0$ (see Eqs (8)-(10) in \cite {kfk}) 
which is proclaimed to be fixed within the factor 1.5 from some theoretical 
arguments about the correlation length of the gluon vacuum expectation value. 
All that concerns the nuclear part of the amplitude for transferred momenta 
$0.05<\vert t\vert <2$ GeV$^2$. Additional parameters have to be ascribed for 
outer intervals. They are related to the Coulomb-nuclear interference region
at very small transferred momenta and to the three-gluon exchange term 
assumed to be relevant at $\vert t\vert >2$ GeV$^2$. Other models
mentioned above use the comparable number of adjustable parameters.
Hard computer work is required to reveal the otherwise hidden
impact of a particular coefficient on the quality of the fit. That is why 
it is desirable to get a simplified model with direct analytical estimates
of this impact and smaller number of parameters. 

\section{The model}

Here, such a model aimed on the rather accurate qualitative description of 
experimental results is presented. Its main outcome contains a single 
parameter only. The proposed model is strongly inspired by the 
phenomenological QCD-motivated kfk-model \cite{kfk, kfk1}, which describes 
experimental data quantitatively in a wide energy interval. That is why
we review first the main findings of the kfk-model. They are at the ground
of the simplified model.

The crucial assumption of the proposed model is the complete neglect 
by the real part of the elastic scattering amplitude $f_R$ at high energies.
Such an assumption can be guessed from the results of the kfk-model.

It was well known from the dispersion relations \cite{dnaz, blo, bloh} that
the real part at high energies is much smaller than the imaginary part $f_I$
for the forward scattering: $f_R(s,t=0)=(0.1 - 0.14)f_I(s,t=0)$. It is 
confirmed by LHC results and satisfied within the kfk-model. Moreover, the real 
part in the kfk-model becomes even much smaller at low transferred momenta 
within the diffraction cone compared to the imaginary part
(see Fig. 3 of \cite{kfk}). It possesses zero inside there in accordance 
with theoretical claims of Refs \cite{mar, mar1}. The integral contribution 
of the real part of the amplitude to the elastic cross section amounts 
to less than 1.5$\%$ since the diffraction cone dominates.
It is demonstrated in the Table II of \cite{kfk1}. The role of the real part 
becomes noticeable for the differential cross section only at its dip where
the imaginary part vanishes as seen from Fig. 4 of \cite{kfk}. 
However the integral contribution from this region is negligible because all
the values at the dip are very low. The position of the dip $t_{dip}$ 
practically coincides with the position of the zero of the imaginary part $t_0$ 
(see Fig. 4 in \cite{kfk1}) because the cross section at the dip is much 
smaller than its values in the diffraction peak. Namely, 
$d\sigma /dt\vert _{dip}/d\sigma /dt\vert _{t=0}\approx 3\cdot 10^{-5}$
at 7 TeV, i.e. the real part at the dip $f_R(t=t_{dip})$ is less 
than $5\cdot 10^{-3}$ of the imaginary part in the diffraction peak 
$f_I(t=0)$.  

These findings validate the neglect by the real part of the amplitude $f_R$ 
in the simplified approach. One can approximate the differential cross section 
by the following expression
\begin{equation}
\frac {d\sigma }{dt}\approx f_I^2
\label{dsdt1}
\end{equation}
neglecting $f^2_R$ compared to $f^2_I$.

The two most typical features of the imaginary part of the amplitude in
the kfk-model are its steep exponential decrease at low transferred momenta
in the so-called diffraction cone and its single zero at some transferred 
momentum. These features can be accounted by the following expression for the 
imaginary part $f_I$ of the nuclear amplitude of the elastic scattering of 
high energy protons used in our simplified model:
\begin{equation}
f_I(s,t)= \frac {\sigma _{tot}(s)}{4\sqrt {\pi}}(1-(t/t_0(s))^2)e^{B(s)t/2}. 
\label{fi}
\end{equation}
The variables $s$ and $t$ are the squared energy and transferred momentum of 
colliding protons in the center-of-mass system $s=4E^2=4(p^2+m^2)$, 
$-t=2p^2(1-\cos \theta)$ at the scattering angle $\theta $. The amplitude is 
normalized according to the optical theorem. Its two typical features are the 
exponential factor with the slope $B$ which governs mainly the 
behavior of the diffraction cone measured at low transferred momenta $t$ and 
the zero at $t=t_0$ which is crucial for the description of the dip/bump region.
Thus there are only two energy-dependent parameters $t_0$ and $B$ in the model.
We consider the total cross section as fixed by the optical theorem at the 
normalization point $t=0$. The formula (\ref{fi}) fits quite well the
kfk-graph for $T_I$ in Fig. 4 of \cite{kfk} in the interval 
$0.05<\vert t\vert <2$ GeV$^2$. The negative term in the brackets steepens
the shape of the diffraction cone that is often approximated by another 
exponent with a larger slope at the end of the cone.

Thus the differential cross section is given by the following expression
\begin{equation}
\frac {d\sigma }{dt}\approx f_I^2=\frac {\sigma ^2 _{tot}(s)}{16\pi}
(1-(t/t_0(s))^2)^2e^{B(s)t}.
\label{dsdt}
\end{equation}
It coincides practically with $d\sigma ^I/dt$ shown in Fig. 4 of \cite{kfk} for 
$\vert t_0\vert =0.4757$ GeV$^2$ and $B\approx B^I=19.90$ GeV$^{-2}$ at 7 TeV
given in the Tables I and II of \cite{kfk1}.
Therefore we do not reproduce their almost identical shapes here.

Surely, our assumption leads to the zero of the differential cross 
section at $t=t_0$ (as in Figs 3, 4 of \cite {kfk, kfk1} for $f_I$ and
$d\sigma ^I/dt$) instead of the dip but rather accurately reproduces 
its behavior in other $t$-regions which are more important for the integral 
contributions. 

The elastic cross section is
\begin{equation}                                                    
\sigma _{el}=\frac {\sigma ^2 _{tot}(s)}{16\pi B}\left (1-\frac {4}{(B t_0)^2}
+\frac {24}{(Bt_0)^4}\right ).
\label{el}
\end{equation}
The structure of the obtained expression is very transparent. The main 
normalization factor $\sigma ^2 _{tot}(s)/16\pi B$ is determined by the
height of the diffraction peak $\sigma _{tot}$ defined by the unitarity 
condition at $t=0$ and its width $B$. The terms in the brackets demonstrate
the suppression of the diffraction peak at its end. Would the real part of the 
amplitude be taken into account, this expression is multiplied by the factor 
$\approx 1+(f_R(s,0)/f_I(s,0))^2\approx 1.02$. To be more precise, the integral
contributions of real and imaginary parts inside the diffraction cone
should be evaluated that shifts the above factor even closer to 1
for the kfk-model because $f_R$ decreases there faster than $f_I$.

According to experimental measurements the ratio of the elastic to total
cross section increases from ISR to LHC up to the value about 1/4. In what 
follows we use the ratio
\begin{equation}
r=\frac {4\sigma _{el}}{\sigma _{tot}}=\frac {\sigma _{tot}(s)}{4\pi B}
\left (1-\frac {4}{(Bt_0)^2}+\frac {24}{(Bt_0)^4}\right ).
\label{rs}
\end{equation}
It is close to 1 at LHC with both factors near 1. The energy dependence of
the first factor is especially important for $r$ in view of the smallness 
of the correction terms in the brackets.

These terms depend on the single dimensionless product 
$B\vert t_0\vert $. They show how deep the zero position $t_0$ penetrates 
inside the diffraction cone at a given energy. This motion of zero is 
often approximated in exponential fits of experimental data on the
differential cross section by the steeper falling
exponent at the end of the diffraction cone. For the present model it is 
taken into account and mimiced by the negative contribution of the terms in the
brackets. They are small at LHC energies because $B\vert t_0\vert \approx 10$ 
there. However the position of the dip (close to $t_0$) seems to move to 
smaller values with energy increase faster than $B$ increases even in the 
range of LHC energies. It is confirmed by the kfk-model (see Table I and
Table II of \cite{kfk1}). The corrections can become larger at higher 
energies. Then the shape of the diffraction cone modifies and can not be 
treated as the exponential one for small enough values of $\vert t_0\vert $.

The same factor determines the dip/bump structure of the differential cross
section beyond the diffraction cone. The bump position is defined by the zero 
of the derivative of $f_I$. The relative shift of the bump 
position $\vert t_b\vert $ to the dip is given by
\begin{equation}
\frac {\vert t_b\vert - \vert t_0\vert}{\vert t_0\vert}\approx 
\frac {2}{B\vert t_0\vert }.
\label{tbt0}
\end{equation}
This ratio depends on the same product $B\vert t_0\vert $. It is about 0.2 
at LHC or $\vert t_b\vert - \vert t_0\vert \approx 0.1$. The estimate 
is a qualitative one. The distance between the dip and the bump
can be somewhat larger because the $t$-dependence of the real part
is important in this interval of the transferred momenta.   

\section{The spatial inelastic profile}

The ratio $r$ is very close to another important characteristics of elastic 
processes
\begin{equation}
\zeta (s)=\frac {1}{2\sqrt {\pi }}\int _0^{\infty}d\vert t\vert f_I(s,t).
\label{zeta1}
\end{equation}
It can be called as the darkness factor because it determines the attenuation
in the spatial profiles of interaction regions for elastic and inelastic 
processes in the impact parameter $b$-presentation\footnote{Let us note that 
the integral from 0 to $\vert t_0\vert $ is positive and that from 
$\vert t_0\vert $ to $\infty $ is negative because $f_I$ changes the sign at 
$t_0$.}. The impact parameter $b$ is determined as the transverse distance 
between the trajectories of the centers of the colliding protons. The knowledge 
of the attenuation in inelasic processes at different impact parameters is 
gained from the unitarity condition which connects elastic and inelastic 
channels of the reaction. Here we consider only the strength of the attenuation 
in central ($b=0$) inelastic collisions referring the readers for the 
detailed description of the unitarity condition and for all-$b$-view to the 
reviews cited above.

The unitarity condition for central head-on collisions with $b=0$ reads 
\cite{ijmp, ufn17}
\begin{equation}
G(s, b=0)=\zeta (s)(2-\zeta (s)),
\label{unit}
\end{equation}         
where $G(s,b)$ is the $b$-profile of inelastic collisions. The darkness of 
central inelastic collisions is complete ($G(s,b=0)=1$) at $\zeta =1$.
Both values are critical ones because any slight decline of $\zeta $ 
from 1 by $\pm \delta $ results in much smaller and always negative decline of 
$G(s,0)$ from 1 by $-\delta ^2$. One gets the complete attenuation at 
$\zeta =1$. The attenuation becomes weaker for any value of $\zeta $ which 
differs from 1. Thus the energy behavior of $\zeta $ determines the deformation
of the inelastic profile with energy.

For the proposed simple model one gets
\begin{equation}
\zeta =\frac {\sigma _{tot}(s)}{4\pi B}\left( 1-\frac {8}{(Bt_0)^2}\right ). 
\label{zmod}
\end{equation}
Here, the correction term in the brackets is somewhat different from those 
in the ratio $r$. However, all of them are small at LHC where 
$B\vert t_0\vert \approx 10$. Our analytical estimates show how severe are the 
requirements upon the accuracy of experimental measurements in the vicinity
of the critical values of $r$ and $\zeta $ close to 1 observed at LHC.
The factor $1+(f_R(s,0)/f_I(s,0))^2\approx 1.02$ should be again kept in mind 
if the real part is accounted.

The ratio $r$ is always larger than $\zeta $:
\begin{equation}
\frac {r}{\zeta }=\frac {1-\frac {4}{(Bt_0)^2}+\frac {24}{(Bt_0)^4}}
{1-\frac {8}{(Bt_0)^2}}\approx 1+\frac {4}{(Bt_0)^2}+\frac {88}{(Bt_0)^4}>1.
\label{r/z}
\end{equation}
Actually, this relation is the main goal of our model. The energy behavior of
the ratio $r/\zeta $ determines the evolution of their relative values. It
depends on a single variable $Bt_0$ only but not on $B$ and $t_0$ separately. 
It is about 10 at LHC with $B\approx 20$ 
GeV$^{-2}$ and $\vert t_0\vert \approx \vert t_{dip}\vert \approx 0.5$ GeV$^2$. 
Thus, $r$ exceeds $\zeta $ by less than 5$\%$. Both of them are close to 1.
However the precision of measurements of $r$ is still not high enough. 

This single variable $B\vert t_0\vert $ can be gained from experimental 
results where the exponential fit of the low-$t$ region is done and the 
position of the dip $t_{dip}\approx t_0$ is determined. Using
this formula (\ref{r/z}) one can easily estimate the accuracy of measurements
which is required for obtaining the accurate enough values. 
It can be used to get the value of $\zeta $ after the ratio
of the elastic and total cross section $r$ is measured precisely enough
and the parameter $B\vert t_0\vert $ is defined. 

The increase of the ratio $r$ from 0.67 at ISR energies to about 1 at LHC 
is directly related to the increase of $\zeta $. Therefore their precise values
in that energy interval are very important. The need in better accuracy of
experimental results is evident. The above discussion of the 
attenuation dependence on the darkness factor shows that the values of $r$
about 1 obtained from experimental data at LHC can be considered as critical 
ones. The accuracy of experimental data at LHC is still not enough to get the 
variables $r$ and $\zeta $ with high enough precision near 1. The desired
accuracy is easily estimated with the help of the formula (\ref{r/z}).

Further behavior of these variables at higher energies is especially crucial.
It is reasonable to assume that the values of $r$ will increase following
the (yet unexplained!) trend at lower energies. The tendencies of $\zeta $ 
to saturate at 1 or increase above 1 at higher energies 
would lead to different predictions about the inelastic profiles with complete 
darkness or decreased attenuation at the center, correspondingly.  In the 
kfk-model asymptotical values of $r$ and $\zeta $ are equal to 1.416 and 1, 
correspondingly. The parameter $B\vert t_0\vert $ should become at least twice 
smaller there as follows from Eq. (\ref{r/z}). Thus one predicts
that the dip must move deeper inside the cone at higher energies.
That is the qualitative feature observable in experiment.

The region of the diffraction cone contributes mostly to both $\zeta $ and $r$
because they are defined as integrals of $f_I$ and $f_I^2$.
To be more definite, the role of the region beyond the cone (at
transferred momenta larger than the dip position) is estimated by integration
 from $\vert t_0\vert $ to infinity. Its contribution $\Delta \zeta $
to $\zeta $ happens to be negligibly small and negative:
\begin{equation}
\Delta \zeta \frac {4\pi B}{\sigma _{tot}}=-\frac {4}{B\vert t_0\vert }\left(1-
\frac {2}{B\vert t_0\vert }\right)e^{-B\vert t_0\vert/2}
\label{t0inf}
\end{equation}
which turns out to be about $-2\cdot 10^{-3}$ at LHC. Therefore the role of 
the tail of the differential cross section is very mild.

The unitarity condition imposes the limit $\zeta \leq 2$. It is required by the 
positivity of the inelastic profile (\ref{unit}). Then there are no inelastic 
processes for central collisions ($G(s,0)=0$ according to Eq. (\ref{unit})). 
It is strange that this limit was called as the "black disk". We prefer to call 
it as TEH - the Toroidal Elastic Hollow \cite{ijmp, ufn17}. The inelastic 
profile acquires the toroidal shape with a hole at the very center which allows
only the elastic scattering in there. In principle, such 
regime is not excluded asymptotically but it is not realized in the kfk-model
where $\zeta $ saturates at 1. It asks for the relation
\begin{equation}
 \sigma _{el}=\sigma _{inel}=\sigma _{tot}/2.
\end{equation}
It is not fulfilled at present energies.
The height of the profile of elastic collisions $\zeta ^2$ at the center $b=0$ 
completely dominates and saturates the total profile $2\zeta $ for $\zeta =2$. 

\section{Discussion and conclusions}

If taken into account, the neglected real part of the elastic amplitude would 
ask for many new parameters to be introduced. We have restricted the
considered range of the transferred momenta to those which provide main
contribution to the integral characteristics $r$ and $\zeta $ described above.
The integral contribution of the real part can be definitely omitted there.

The special problem to be discussed is the energy behavior of $\zeta $. It is 
important that the value of $\zeta $ in the kfk-model starts increasing at ISR 
and approaches 1 at LHC energies being below 1 by several percents only. 
At the same time, if the steady increase of the share of elastic processes from 
ISR to LHC persists at higher energies with $r$ becoming larger than 1, one 
should consider the intriguing possibility that $\zeta $ will also exceed 1. 
Surely, that can happen only if the ratio of the total cross section to 
$4\pi B$ becomes noticeably larger than 1. It increased from about 0.67 at 
ISR-energies to 1.02$\pm$0.04 at LHC. Actually, the experimental 
values of $r$ at LHC energies range from 1.01$\pm$0.06 \cite{totem1} 
to 1.06$\pm$0.06 \cite{totem4}. The factor in brackets in Eq.(\ref{rs}) is about 
0.96. Thus the situation at LHC energies is critical in the sense that all 
estimates of $r$ and $\zeta $ are near 1 within the accuracy of experimental 
data. The further insight into proton collisions can be gained if more precise 
data on elastic scattering at LHC will be obtained.
The accuracy of measurements of $\sigma _{tot}, B$ and $t_{dip}$ is decisive.

In the case of the further noticeable increase of $\sigma _{tot}/4\pi B$ at 
higher energies the attenuation for central inelastic collisions can become 
incomplete ($G(s,b=0)<1$) after passing through its completeness at LHC. 
However if the value of $\zeta $ tends to 1 asymptotically, no such peculiar 
behavior will be observed. $G(s,b=0)$ will tend to 1. The last
possibility looks quite probable both from our intuitive expectations and from
conclusions of the kfk-model. 

In conclusion, we have proposed the simple model of elastic scattering of 
high energy protons which admits analytic calculations and easy estimates 
with the help of a single parameter related to experimentally measurable
characteristics. The accuracy of experimental measurements is directly
translated into the required precision of the estimates of its value. The
analytic expressions allow experimentalists to find the necessary demands upon 
 the accuracy of measurements by direct estimation of the product of the cone 
 slope $B$ and the dip position $t_{dip}\approx t_0$ by providing the
accurate values of $r$ and $\zeta $. That is especially important in the LHC 
energy range where both $r$ and $\zeta $ reach their critical values close to 1.

What concerns the further perspective, one can state that there is yet no 
consensus about possibilities for energy behavior of the share of elastic 
processes $r$ and of $\zeta $ at higher energies. Their asymptotical 
approach to 1 or increase above 1 would tell us not only about elastic 
scattering but reveal interesting features of inelastic collisions as well.

\medskip

{\bf Acknowledgments}

\medskip 
 
I am grateful for support by the RAS-CERN 
program and the Competitiveness Program of NRNU "MEPhI" ( M.H.U.).

\end{document}